\documentclass[prl,twocolumn,showpacs,amsmath,superscriptaddress,psfig,amssymb]{revtex4}
\usepackage{amssymb}

\usepackage{graphicx}
\usepackage{dcolumn}
\usepackage{bm}

\begin{document}


\title{Multiband superconductivity in PrPt$_{4}$Ge$_{12}$ single crystals}

\author{J. L. Zhang}
\affiliation{Department of Physics and Center for Correlated Matter,
Zhejiang University, Hangzhou, Zhejiang 310027, China}
\author{Y. Chen }
\affiliation{Department of Physics and Center for Correlated Matter, Zhejiang University, Hangzhou,
Zhejiang 310027, China}
\author{L. Jiao}
\affiliation{Department of Physics and Center for Correlated Matter, Zhejiang University, Hangzhou,
Zhejiang 310027, China}
\author{R. Gumeniuk}
\affiliation{Max-Planck-Institut f$\ddot{u}$r Chemische Physik fester Stoffe,
N$\ddot{o}$thnitzer Str. 40, 01187 Dresden, Germany}
\author{M. Nicklas}
\affiliation{Max-Planck-Institut f$\ddot{u}$r Chemische Physik fester Stoffe,
N$\ddot{o}$thnitzer Str. 40, 01187 Dresden, Germany}
\author{Y. H. Chen }
\affiliation{Department of Physics and Center for Correlated Matter, Zhejiang University, Hangzhou,
Zhejiang 310027, China}
\author{L. Yang}
\affiliation{Department of Physics and Center for Correlated Matter, Zhejiang University, Hangzhou,
Zhejiang 310027, China}
\author{B. H. Fu}
\affiliation{Department of Physics and Center for Correlated Matter, Zhejiang University, Hangzhou,
Zhejiang 310027, China}
\author{W. Schnelle}
\affiliation{Max-Planck-Institut f$\ddot{u}$r Chemische Physik fester Stoffe,
N$\ddot{o}$thnitzer Str. 40, 01187 Dresden, Germany}
\author{H. Rosner}
\affiliation{Max-Planck-Institut f$\ddot{u}$r Chemische Physik fester Stoffe,
N$\ddot{o}$thnitzer Str. 40, 01187 Dresden, Germany}
\author{A. Leithe-Jasper}
\affiliation{Max-Planck-Institut f$\ddot{u}$r Chemische Physik fester Stoffe,
N$\ddot{o}$thnitzer Str. 40, 01187 Dresden, Germany}
\author{Y. Grin}
\affiliation{Max-Planck-Institut f$\ddot{u}$r Chemische Physik fester Stoffe,
N$\ddot{o}$thnitzer Str. 40, 01187 Dresden, Germany}
\author{F. Steglich }
\affiliation{Max-Planck-Institut f$\ddot{u}$r Chemische Physik fester Stoffe,
N$\ddot{o}$thnitzer Str. 40, 01187 Dresden, Germany}
\author{H. Q. Yuan}
\email{hqyuan@zju.edu.cn}
\affiliation{Department of Physics and Center for Correlated Matter, Zhejiang University, Hangzhou,
Zhejiang 310027, China}

\date{\today}

\begin{abstract}

We report measurements of the London penetration depth
$\Delta\lambda(T)$ and the electronic specific heat $C_e(T)$ on
high-quality single crystals of the filled-skutterudite
superconductor PrPt$_4$Ge$_{12}$ ($T_c\simeq$8K). Both quantities
show a weak temperature dependence at $T\ll T_c$, following
$\Delta\lambda\sim T^n$ ($n\simeq 3.2$) and $C_e/T\sim T^{2.8}$. Such temperature
dependences deviate from both conventional $s$-wave type and nodal
superconductivity. A detailed analysis indicates that the superfluid
density $\rho_s(T)$, derived from the penetration depth, as well as
the electronic specific heat can be consistently described in terms
of a two-gap model, providing strong evidence of multiband
superconductivity for PrPt$_4$Ge$_{12}$.

\end{abstract}


\maketitle

The filled-skutterudite compounds $MT_4X_{12}$ ($M$=rare-earth or
alkaline-earth metals, $T$=Fe, Ru, Os, and $X$=P, As, Sb)
demonstrate remarkably rich physical properties \cite{Sales}.
Particular attention has been paid to superconductivity (SC)
observed in the Pr-based compounds. For example, PrOs$_4$Sb$_{12}$
is a heavy-fermion superconductor with $T_c=1.85$ K \cite{Bauer}.
Electrical-quadrupole, rather than magnetic-dipole, fluctuations are
believed to mediate the Cooper pairs in this compound, which is
unique among heavy fermion superconductors. The superconducting
order parameter of PrOs$_4$Sb$_{12}$ remains controversial: nodal SC
\cite{Izawa, Chia} as well as $s$-wave SC \cite{MacLaughlin} were
proposed. More recent experiments seem to support a scenario of
multiband SC \cite{Seyfarth06, Hill}. On the other hand, the
isostructural compounds PrRu$_4$Sb$_{12}$ and PrRu$_4$As$_{12}$
appear to be $s$-wave superconductors \cite{Takeda, Namiki}.

Recently, a series of new skutterudite superconductors with a
germanium-platinum framework, i.e., $M$Pt$_4$Ge$_{12}$ ($M$=Sr, Ba,
La, Pr), were successfully synthesized \cite{Bauer2007, Gumeniuk}.
Among all the Pr-filled variants, PrPt$_4$Ge$_{12}$ shows an
unexpectedly high transition temperature of $T_c$=7.9 K
\cite{Gumeniuk}. The Sommerfeld coefficient of PrPt$_4$Ge$_{12}$
($\gamma_n=76$ mJ/mol K$^{2}$) \cite{Maisuradze08} is comparable to
that of PrRu$_4$Sb$_{12}$ \cite{Takeda} and PrRu$_4$As$_{12}$
\cite{Namiki}, but much smaller than that of PrOs$_4$Sb$_{12}$
\cite{Bauer}. Furthermore, the crystalline electric field (CEF)
splitting of the $J=2$ Hund's rule multiplet of Pr$^{3+}$ between
the ground state and the first excited state is rather different
among these Pr-based superconductors, e.g.,
$\Delta_{\textrm{CEF}}=7$ K for PrOs$_4$Sb$_{12}$ \cite{Maple} and
$\Delta_{\textrm{CEF}}=$130 K in PrPt$_4$Ge$_{12}$ \cite{Gumeniuk,
Toda}. It is, therefore, of great interest to systematically compare
the superconducting properties of these materials, which may help to
elucidate their pairing mechanisms. Similar to PrOs$_4$Sb$_{12}$,
previous studies on polycrystalline samples of PrPt$_4$Ge$_{12}$
showed controversial results. Measurements of the specific heat and
muon-spin rotation ($\mu$SR) suggest the possible existence of point
nodes in the superconducting gap \cite{Maisuradze08}; zero-field
$\mu$SR also provides evidence of time reversal symmetry breaking
below $T_c$ \cite{Maisuradze09}, similar to what was observed for
PrOs$_4$Sb$_1$$_2$ \cite{Aoki}. However, $^7$$^3$Ge nuclear
quadrupole resonance (NQR) experiments display a pronounced
coherence peak in the spin-lattice relaxation rate $1/T_1$ at
temperatures just below $T_c$, suggesting $s$-wave SC
\cite{Kanetake}. Very recently, a possible scenario of multiband SC
was proposed for PrPt$_4$Ge$_{12}$, based on the analysis of the
critical fields \cite{Sharath} as well as photoemission spectroscopy
\cite{Nakamura}. However, these experiments were performed on
polycrystalline samples at relatively high temperatures, which could
not make a clear assertion on the gap symmetry. The reasons
underlying such discrepancies of the gap structure in
PrPt$_4$Ge$_{12}$ are not yet clear, and further measurements, in
particular those based on high-quality single crystals, are badly
needed.

In this Letter, we probe the superconducting gap symmetry of
PrPt$_{4}$Ge$_{12}$ by measuring the London penetration depth
$\Delta\lambda(T)$ and the specific heat $C_p(T)$ of high-quality
single crystals. Precise measurements of the penetration
depth changes at low temperatures show $\Delta\lambda\sim T^n$ with $n\simeq
3.2$, indicating that PrPt$_4$Ge$_{12}$ is actually neither a simple
BCS nor a nodal superconductor. A detailed analysis of the
superfluid density $\rho_s(T)$, converted from $\lambda(T)$,
and the electronic specific heat $C_e(T)$ provide strong evidence of
two-band SC for PrPt$_4$Ge$_{12}$.

\begin{figure}[b]\centering
 \includegraphics[width=7.5cm]{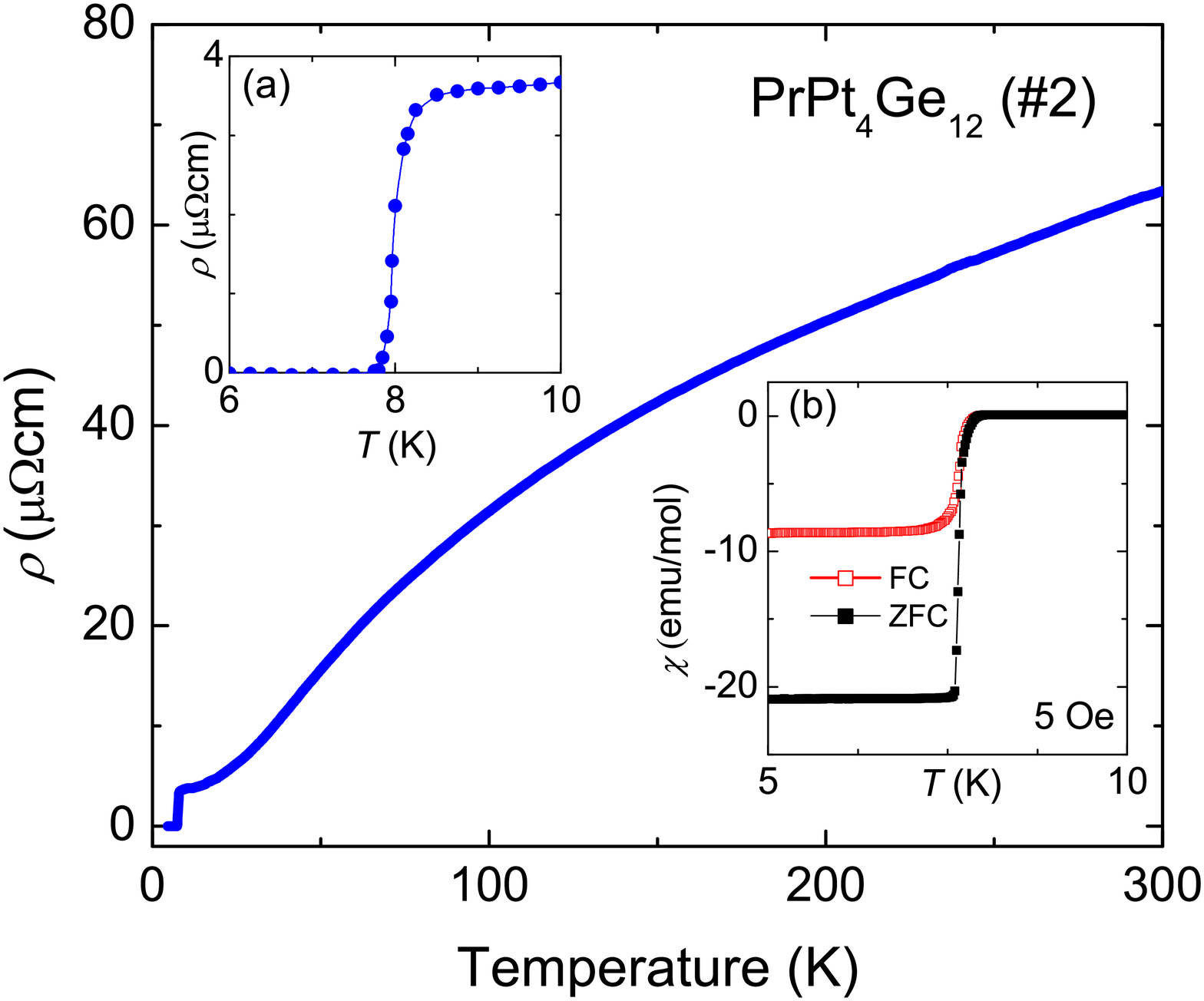}
\caption{Temperature dependence of
the electrical resistivity $\rho(T)$ for PrPt$_{4}$Ge$_{12}$.
Insets show the superconducting transitions in the electrical resistivity $\rho(T)$ (a) and magnetic susceptibility $\chi(T)$ (b), respectively.}\label{fig.2}
\end{figure}

High-quality single crystals of PrPt$_4$Ge$_{12}$ were synthesized
by using multi-step thermal treatments \cite{Methods}. Powder X-ray
diffraction indicates the presence of a small amount of foreign
phases. Energy-dispersive X-ray (EDX) analysis confirms that all the
crystals have a stoichiometric composition and the impurity phases,
mainly PtGe$_2$ and free Ge, are located at the crystal surfaces
\cite{Methods}. In our measurements, the crystals were mechanically
polished to get rid of these surface contaminations. Precise
measurements of the resonant frequency shift $\Delta f(T)$ were
performed by utilizing a tunnel diode oscillator (TDO) based,
self-inductance method at an operating frequency of 7 MHz down to
about 0.5K in a $^{3}$He cryostat \cite{Van Degrift}. The change of
the penetration depth is proportional to the resonant frequency
shift, i.e., $\Delta\lambda(T)=G\Delta f(T)$, where $G$ is solely
determined by the sample and coil geometries \cite{Chia}. In this
context, $\Delta\lambda(T)$ is extrapolated to zero at $T=0$ by
polynomial regression, i.e.,
$\Delta\lambda(T)=\lambda(T)-\lambda_0$. Here the value of
zero-temperature penetration depth, $\lambda_0$=114 nm, was adopted
from previous $\mu$SR experiments \cite{Maisuradze08}. Measurements
of the magnetic susceptibility were carried out in a SQUID
magnetometer (Quantum Design), and the heat capacity was measured in
a $^3$He cryostat, using a relaxation method.

\begin{figure}[b]\centering
 \includegraphics[width=7.5cm]{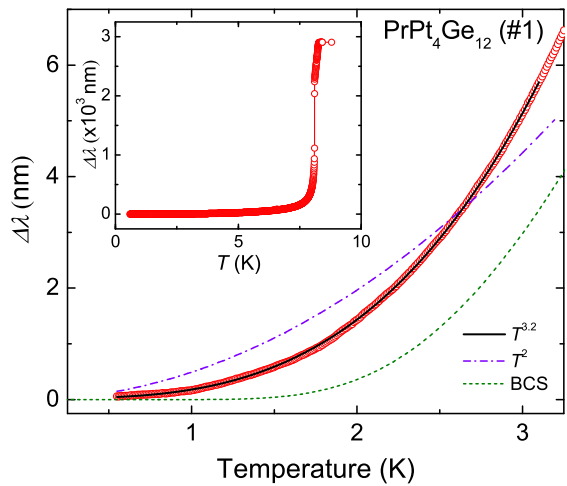}
\caption{Temperature dependence of the penetration depth
$\Delta\lambda$(T) for PrPt$_{4}$Ge$_{12}$. Solid, dash-dotted and
dashed lines represent the fit of $\Delta\lambda\sim T^{3.2}$,
$\Delta\lambda\sim T^{2}$ (point node) and single-gap BCS model,
respectively. Inset shows $\Delta\lambda(T)$ over the entire
temperature range.}\label{fig.3}
\end{figure}

Figure 1 presents the electrical resistivity $\rho(T)$ for
PrPt$_{4}$Ge$_{12}$, which shows an S-shape behavior upon cooling
down from room temperature, as often observed in $d$-band materials.
A sharp superconducting transition, evidenced from both the
electrical resistivity $\rho(T)$ (inset (a)) and the magnetic
susceptibility $\chi(T)$ (inset (b)), together with a large
resistivity ratio ($\rho$(300K)/$\rho$(8K)=19) confirm the high
quality of our single crystals. Furthermore, the superconducting
transition temperatures $T_c$, determined from the zero resistivity
and the onset of the magnetic susceptibility, are nearly the same
($T_c\simeq$7.8 K), proving good homogeneity of the samples.

The inset of Fig. 2 shows the change of the penetration depth
$\Delta\lambda(T)$ up to 9K, which reveals a sharp superconducting
transition at $T_c\simeq8.1$ K. Here $G$=2.13\AA/Hz. It is noted that we have measured the penetration depth for several samples and the data are well reproducible. The values of $T_c$, derived for different samples by distinct methods, are nearly same too. In the main plot of Fig. 2, we presents
$\Delta\lambda(T)$ at low temperatures, together
with the fits of various models to the data.
Obviously, the standard BCS model can not describe
the experimental data. Moreover, the penetration depth
$\Delta\lambda(T)$ deviates also from that of nodal SC, for which a
linear and quadratic temperature dependence is expected for the
case of line and point nodes, respectively. Instead, a power law of
$\Delta\lambda\sim T^{n}$ ($n\simeq 3.2$) presents a reasonable fit to
the experimental data. An enhanced power-law exponent $n$, e.g., a
quadratic temperature dependence in $d$-wave superconductors, may
arise from nonlocal effects or impurity scattering
\cite{Hirschfeld}. However, such possibilities are excluded for
PrPt$_{4}$Ge$_{12}$ because both the penetration depth
($\lambda_0=$114 nm) \cite{Maisuradze08} and the mean free path
($l=103$ nm) are much larger than the coherence length
($\xi_{0}=13.5$ nm) \cite{Maisuradze08}, implying that the samples
are in the clean and local limit. Here we estimate the mean free
path from
$l=[\frac{\xi_0^{-2}-1.6\times10^{12}\rho_0\gamma_nT_c}{1.8\times
10^{24}(\rho_0\gamma_nT_c)^2}]^{1/2}$ \cite{Orlando}, where
$\rho_0$, $\xi_0$ and $\gamma_n$ represent the electrical
resistivity at $T_c$=7.8K ($\rho_0=3.5\times10^{-6}$ $\Omega$cm),
the aforementioned coherence length and the Sommerfeld coefficient
at $T_c$ ($\gamma_n=$1795 erg cm$^{-3}$K$^{-2}$), respectively. On
the other hand, multiband effects may also give rise to power-law-like
behavior at low temperatures with a large exponent $n$, which will be further elucidated by
the analysis of both the superfluid density and specific heat.

The superfluid density $\rho_s(T)$ provides an important
characterization of the superconducting gap symmetry. Fig. 3 shows
the temperature dependence of the normalized superfluid density
$\rho_s(T)$ for PrPt$_4$Ge$_{12}$ (circles), which is calculated by
$\rho_{s}=[\lambda_0/\lambda(T)]^{2}$. For comparison, in Fig.3 we
also include the superfluid density from the $\mu$SR results
determined on polycrystalline samples (diamonds)
\cite{Maisuradze08}. Obviously, these two data sets are quite
compatible, although the $\mu$SR data are more scattered, with a
poor resolution when compared with the TDO results. This allows to
probe the gap structure of PrPt$_4$Ge$_{12}$ in a much more precise
way than before.

\begin{figure}[b]\centering
 \includegraphics[width=7.5cm]{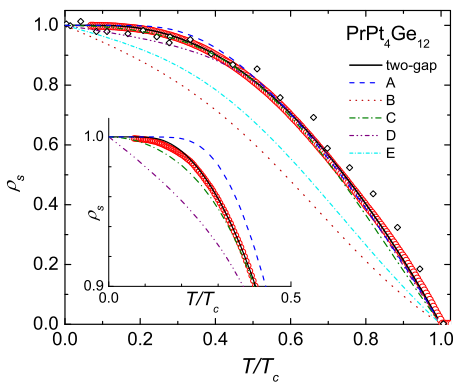}
\caption{Superfluid density $\rho_s(T)$ versus normalized
temperature $T/T_c$. Inset expands the low temperature region.
Circle ($\circ$) and diamond ($\diamond$) dispaly the experimental
data derived from TDO measurements (this study) and $\mu$SR
experiments (from Ref.\cite{Maisuradze08}), respectively. Lines
shows theoretical fits of various gap functions as listed in Table
1. }\label{fig.4}
\end{figure}

The superfluid density can be calculated by:
\begin{equation}
\rho_s(T)=1+2\langle\int^{\infty}_{0}\frac{\partial f}{\partial E}
\frac{E}{\sqrt{E^{2}-\Delta_k^{2}(T)}}dE\rangle_{\textrm{FS}},
\end{equation}
where $f=(e^{\sqrt{{E}^2+\Delta_k^2(T)}/k_BT}+1)^{-1}$ is the Fermi
distribution function and $\langle\ldots\rangle_{\textrm{FS}}$
denotes the average over the Fermi surface. For the temperature
dependence of the energy gap, we take
$\Delta(T)=\Delta_{0}\tanh\frac{\pi
k_BT_{c}}{\Delta_{0}}[\frac{2}{3}\frac{\Delta
C_e}{\gamma_nT_c}(\frac{T}{T_{c}}-1)]^{0.5}$ \cite{Prozorov 11}. Here
$\Delta C_e$ is the specific heat jump at $T_c$. Note that Eq. 1 is
applicable for various gap functions $\Delta_k$ (=$\Delta(\theta,
\phi)$) in the pure/local limit. Given a gap function
$\Delta(\theta, \phi)$, then one can fit it to the experimental
data. Here $\theta$ and $\phi$ denote the angles away from the
z-axis and x-axis in k-space, respectively. In this analysis, the
zero-temperature gap amplitude, $\Delta_0$, is the sole fitting
parameter.

Possible symmetries of the order parameter have been theoretically
investigated for the skutterudite superconductors with tetrahedral
point group symmetry ($T_h$) \cite{Sergienko}. Various gap functions
$\Delta(\theta,\phi)$, restrained by the crystal symmetry, have been
adopted to fit the superfluid density $\rho_s(T)$ of
PrOs$_4$Sb$_{12}$ \cite{Chia}. In this context, we apply a similar
analysis to the experimentally obtained $\rho_s(T)$ data of
PrPt$_4$Ge$_{12}$. Fig. 3 presents the fits of different gap
functions allowed by the crystal symmetry; the derived fitting
parameters of $\Delta_0$ are summarized in Table 1. Apparently, the
gap functions B and E cannot reproduce the experimental data. On the
other hand, the fits of functions A, C and D are close to the
experimental data, but show significant deviations at low
temperatures (inset of Fig.3). We note that models C and D, both
having point nodes in the superconducting gap, were previously
assumed to present a good fit to the $\mu$SR data
\cite{Maisuradze09}, which are rather scattered at low temperature. The more precise measurements
of the penetration depth, $\Delta\lambda (T)$, and the corresponding
superfluid density, $\rho_s(T)$, indicate that the conventional
one-gap BCS model as well as the nodal-gap model D provide a poor
fit to the low temperature data. The nodal-gap model C can fit the
TDO data relatively well, but significant deviations remain below
0.4$T/T_c$. Instead, the two-gap BCS model gives the best fit to our
experimental data.

In the case of two-gap BCS superconductors, the superfluid density can be extended to
the following linear combination\cite{Manzano}:
\begin{equation}
\widetilde{\rho}_{s}(T)=x\rho_s(\Delta_0^1,T)+(1-x)\rho_s(\Delta_{0}^2,T),
\end{equation}
where $\Delta_0^i (i=1,2)$ represent the size of two isotropic gaps
at zero temperature, and $x$ is the relative weight of the
contributions from $\Delta_0^1$. As shown in Fig. 3, the two-gap BCS
model nicely fits the experimental data of PrPt$_4$Ge$_{12}$ over
the entire temperature region. The so derived parameters of
$\Delta_0^1=0.8k_BT_c$, $\Delta_0^2=2.0k_BT_c$ and $x$=0.15 meet the
theoretical constraints that one gap is larger than the BCS value
and the other one is smaller \cite{Kresin}, as demonstrated in the
prototype two-gap BCS superconductor MgB$_2$ \cite{Manzano}.

\begin{table}[tbp]
\centering \caption{Summary of various gap functions and so derived
fitting parameters $\Delta_0$.}
\begin{tabular}{lccc} \hline model &Gap function
$\Delta(\theta,\phi)$ &$\Delta_0/k_{B}T_{c}$ \\ \hline
A &$\Delta_0$ &1.76 \\
B &$|\Delta_0\sin\theta\sin\phi|$ &4 \\
C &$|\Delta_0\sin\theta|$ &2.7 \\
D &$\Delta_0(1-\sin^{4}\theta\cos^{4}\theta)$ &2.9 \\
E &$\Delta_0(1-(\sin^{4}\phi+\cos^{4}\phi)\sin^{4}\theta)$
&3.2\\\hline
\end{tabular}
\end{table}

\begin{figure}[b]\centering
 \includegraphics[width=7.5cm]{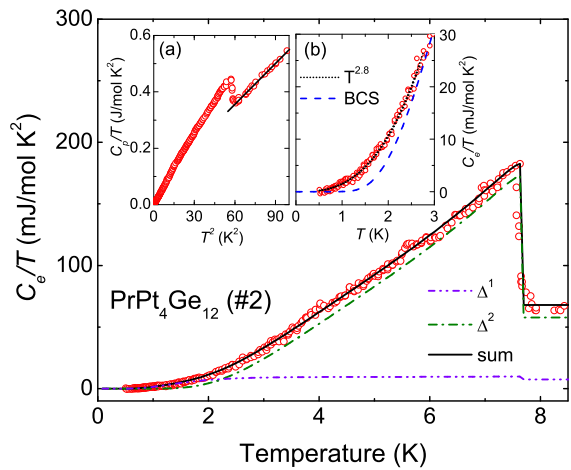}
\caption{Temperature dependence of the electronic specific heat
$C_e(T)/T$ of a PrPt$_{4}$Ge$_{12}$ single crystal at zero field.
Dash dotted lines (green and blue) and solid line show the
individual and total contributions of the two gaps to the specific
heat $C_e(T)/T$, respectively. Inset (a): total specific heat,
$C_p(T)/T$, plotted as a function of $T^2$. Inset (b): electronic
specific heat $C_e(T)/T$ at low temperatures, fitted by $C_e/T\sim
T^{2.8}$ (dotted line). Dashed line refers to the standard BCS
model. }\label{fig.5}
\end{figure}

In Fig. 4, we present the low-temperature specific heat, $C_p(T)$, of
a PrPt$_4$Ge$_{12}$ single crystal. A sharp superconducting
transition is observed at $T_{c}=7.7$ K, being close to that
determined from other experiments. The specific heat data above
$T_c$ can be fitted by a polynomial expansion $C_p(T)= \gamma_nT+
\beta T^3$. Here $C_e=\gamma_nT$ and $C_{ph}=\beta T^3$ denote the
electronic and phonon contributions, respectively. This yields the
Sommerfeld coefficient in the normal state, $\gamma_n=69$ mJ/mol
K$^2$, and the Debye temperature $\Theta_D=$190 K, which are close to
those found in case of polycrystalline samples \cite{Maisuradze08}. For
polycrystals \cite{Maisuradze08}, a pronounced upturn
was previously reported in the low-temperature specific heat
$C_e(T)/T$. Similar specific heat anomalies were also observed in
some as-grown single crystals. A careful examination showed that
such an upturn in $C_e(T)/T$ has to be attributed to a nuclear
Schottky anomaly caused by the Pr-containing surface contaminations
\cite{Methods}. Indeed, the specific heat anomaly disappears for the
polished single crystal as shown in Fig.4(b), allowing us to
accurately analyze its low temperature behavior.

The electronic specific heat of PrPt$_4$Ge$_{12}$,
obtained after subtracting the phonon contributions, is presented in
the main part and inset (b) of Fig. 4 as $C_e/T$ verse $T$, together with the fits of
various models. As shown in the inset (b), the data can be well
described by a power law, $C_e/T\sim T^{2.8}$. This behavior
deviates from the quadratic temperature dependence of $C_e(T)/T$
reported in Ref. \cite{Maisuradze08}. The discrepancy is likely to
result from the nuclear Schottky anomaly of the polycrystalline
samples discussed before. With the previous data
\cite{Maisuradze08}, a proper subtraction of this Schottky anomaly
is difficult and, therefore, deviations from the true specific-heat
behavior become likely at low temperatures. Furthermore, the
standard BCS-model is not sufficient to fit the experimental data
(Fig. 4(b)), while the two-gap BCS model presents the best fit to the
$C_e(T)/T$ data (main figure). According to the phenomenological
two-gap BCS model, the heat capacity is taken as the sum of
contributions from the two bands, each one following the BCS-type
temperature dependence \cite{Bouquet}. In the main panel of Fig. 4,
we plot the contributions from the two superconducting gaps,
$\Delta_0^1=0.8k_B T_c$ and $\Delta_0^2=2.0k_B T_c$, as well as
their sum (solid line). The weight contributed from the first gap,
$\Delta_0^1$, is $x$=0.12. All these fitting parameters are
remarkably consistent with those obtained from the superfluid
density $\rho_s(T)$, providing strong evidence of two-gap SC for
PrPt$_4$Ge$_{12}$.

Evidence of BCS-like SC, including two-gap type, has been observed
in several skutterudite compounds. For example, PrRu$_4$Sb$_{12}$
\cite{Takeda}, PrRu$_4$As$_{12}$ \cite{Namiki} and their non-4$f$
counterparts \cite{Maisuradze08} are believed to be $s$-wave
superconductors. Recent measurements indicate that PrOs$_4$Sb$_{12}$
is an extreme two-band superconductor \cite{Seyfarth06, Hill}; here
energy nodes were assumed to exist in the small gap, and the
isotropic large gap dominates the superconducting properties near
$T_c$ \cite{Hill}, or when a sufficiently large magnetic field is
applied \cite{Shu}. Two-gap BCS SC was also proposed for both
PrRu$_4$Sb$_{12}$ \cite{Hill} and LaOs$_4$Sb$_{12}$ \cite
{Chia2012}, the latter one suggesting that 4$f$-electrons are not
the origin of multiband SC. In PrPt$_4$Ge$_{12}$, band structure
calculations indicate an only minor contribution of the
4$f$-electrons to the density of states at the Fermi energy,
suggesting that the 4$f$-electrons may not be playing a significant
role on SC in this compound either \cite{Gumeniuk}. Indeed, the
thermodynamic properties and low-lying CEF scheme of
PrPt$_4$Ge$_{12}$ seem to be rather different from those of the
heavy fermion compound PrOs$_4$Sb$_{12}$, but resemble other
skutterudite compounds \cite{Takeda, Namiki}. Indications of two-gap
SC for PrPt$_4$Ge$_{12}$ were also inferred from recent measurements
of the upper and lower critical fields \cite{Sharath} and
photoemission spectroscopy \cite{Nakamura}. Furthermore, multiband
SC is compatible with the observations of a coherence peak in the
NQR measurements just below $T_c$ \cite{Kanetake}. Such a multi-gap
structure seems to be characteristic for the skutterudite
superconductors; the small gap, either with or without nodes, is
rather subtle and can be easily destroyed by external effects, e.g.,
a magnetic field, so that the large gap is predominant. Recent $\mu$SR
measurements performed on polycrystalline samples of PrPt$_4$Ge$_{12}$ showed evidence of time-reversal symmetry breaking \cite{Maisuradze09}. To confirm it and check the possible existence of nodes in the small gap of
PrPt$_4$Ge$_{12}$, it would be important to repeat these measurements with high-quality single crystals. Detailed calculations of its
electronic structure are also highly desirable in order to further elucidate the multiband
structure in PrPt$_4$Ge$_{12}$. Moreover, comparative
studies of the Pr-based skutterudites and the non-$f$ electron
isostructural compounds, e.g., $M$Pt$_4$Ge$_{12}$ ($M$ = Sr, Ba and
La), are necessary to reveal the potential role of
$f$-electrons on SC.

In summary, we have studied the superconducting order parameter of
PrPt$_{4}$Ge$_{12}$ by measuring the penetration depth
$\Delta\lambda(T)$ and specific heat $C_p(T)$ on high-quality single
crystals. For $T\ll T_c$, both quantities demonstrate a weak
temperature dependence and can be fitted by a power-law behavior
with a large exponent, i.e., $\Delta\lambda\sim T^{3.2}$ and $C_e/T\sim
T^{2.8}$, which is inconsistent with both a single-gap BCS model and
nodal-gap SC. Instead, we can describe the superfluid density
$\rho_s(T)$ and the electronic specific heat $C_e(T)$ in terms of a
phenomenological two-gap BCS model with consistent gap parameters of
$\Delta_0^1=0.8k_BT_c$, $\Delta_0^2=2.0k_BT_c$ and $x=0.12\sim
0.15$, the weight contributed by the small gap. These findings have
elucidated the controversial results found in the literature and
provide unambiguous evidence of multiband SC for PrPt$_4$Ge$_{12}$.

\acknowledgments

We are grateful to E. Rosseeva, Y. Kohama and C. T. van Degrift for providing experimental assistance. This work was supported by the
National Basic Research Program of China (NOs. 2009CB929104 and 2011CBA00103), NSFC (NOs. 10934005 and 11174245), Zhejiang Provincial
Natural Science Foundation of China, the Fundamental Research Funds for the Central Universities and the Max-Planck Society under the auspices of the Max-Planck Partner Group of the MPI for Chemical Physics of Solids, Dresden.

\end{document}